\documentstyle[preprint,aps]{revtex}
\begin{document}
\draft
\preprint{\vbox{
\hbox{IFT-P.031/95}
\hbox{hep-ph/9506059}
\hbox{June 1995}
}}
\title{Concerning the vacuum velocity of gravitational waves}
\author{Vicente Pleitez }
\address{
Instituto de F\'\i sica Te\'orica\\
Universidade Estadual Paulista\\
Rua Pamplona, 145\\
01405-900-- S\~ao Paulo, SP\\
Brazil}
\maketitle
\begin{abstract}
It is pointed out that if gravitational interactions among ordinary bodies
propagate in extra
space-time dimensions the velocity of gravitational waves in vacuum could
be different from the speed of light $c$.
\end{abstract}
\vskip .5cm
PACS numbers: 04.50.+h; 
 04.30.Nk 
\newpage
One of the main issues to be set by the detection programs of gravitational
waves
(like LIGO and VIRGO projects) is the vacuum velocity of gravitacional waves
($c^*$), and also their polarization.
In the context of general relativity this velocity is the same
as the light velocity ($c$). For instance, if this velocity were found to be
smaller than $c$ it would imply, at first sight, that gravitons are massive.
However,
it has been shown, by Van Dam and Veltman~\cite{gra1}, through the
measurement of the perihelion movement of Mercury and the bending of a ray
of light passing near the Sun, that there is a discrete difference between
the zero-mass
theories and the very small but non-zero mass theories. Hence, consistency with
the data imply that the graviton
must be rigorously massless.
As a consequence, if the measured velocity of the
gravitational waves were different from the velocity of light, a problem
would come out for theorists.

In this work we shall put forward a non-orthodox idea: gravitons
are massless and although they are generated by ordinary matter, they
propagate through a different space-time (which we will
call $x^*$-world). In this world the speed limit in vacuum is $c^*$, and
it is, in principle, different from $c$, the speed limit in the $x$-world.
Connection between both $x$- and $x^*$-worlds is possible by assuming
appropriate transformation properties for matter and radiation fields under
an extended Lorentz symmetry~\cite{locos}.

In order to implement this idea, let us assume that the Lorentz symmetry of
an eight-dimensional world is $L\otimes L^*$
being $L$ and $L^*$ Lorentz groups with different limit velocities $c$ and
$c^*$, respectively.
We denote $x$ and $x^*$ the space-time
which transform as $({\bf4},{\bf1})$ and $({\bf1},{\bf4})$ under
$L\otimes L^*$  respectively. Our own world is identified with the $x$-world
and the transformation properties of all known spin
$0,1/2,1$ fields are defined with respect to the $L$ group but all of them
transform as scalars under $L^*$.
{}From the point of view of quantum field theory, gravitation
is caused by the exchange of a particle of spin-2~\cite{gra2}.
Then, the graviton is assumed to be described by a second rank symmetric
tensor in the Minkowski space related with $L^*$ but it is a scalar under $L$.

The interaction between two material objects (denoted by $a$ and $b$) caused
by the exchange of a
graviton in the Born approximation is~\cite{gra1}
\begin{equation}
g^2\,T^a_{\mu^*\nu^*}\,P_{\mu^*\nu^*\alpha^*\beta^*}
\,T^b_{\alpha^*\beta^*},
\label{int}
\end{equation}
where $P_{\mu^*\nu^*\alpha^*\beta^*}$ is the usual graviton propagator
{}~\cite{gra1,gra2} but now the propagation occurs in the $x^*$-world. The
traceless tensors $T$ in Eq.~(\ref{int}) can be formed with the scalar
part (under $L^*$) of fields with different transformation
properties under $L$.

The coupling constant $g$ is fixed by the requirement
that Eq.~(\ref{int}) contains the Newton law for non-relativistic bodies.
Usually the coupling with dimension of $\mbox{(mass)}^{-2}$ is written as
$g^2\rightarrow G_N/\hbar c$. In our context, however, we have
\begin{equation}
g^2\to g^{*2}=F^2\,\frac{c}{c^*}\,\frac{G_N}{\hbar c}\equiv
\frac{G^{\rm eff}_N}{\hbar c}.
\label{g}
\end{equation}

The dimensionless factor $F^2$ appearing in Eq.~(\ref{g}) arises as follows.
The interaction in Eq.~(\ref{int}) occurs in the $x^*$ world, i.e., the
tensors $T^{a,b}_{\mu^*\nu^*}$ are built  with the
usual fields which transform as scalars under $L^*$. However, since we are
interested in the gravitational effects
observed in the $x$-world where we live, we can integrate over a finite volume
  of the $x$-world,
so it is possible to calculate the dimensionless $F$. There is a
factor $F$ in each vertex. Its value, in principle, depends on the process
under consideration since the volume of integration may differ from one
process to another. Similarly, a factor ${\cal F}$ will appear in the usual
interactions among fields in the $x$-world. However, in this case we can
integrate over the whole volume
of the $x^*$-world. Thus, in this case ${\cal F}=1$ and we obtain the usual
interaction, say, between spin-$\frac{1}{2}$ and photon fields. The
interactions among gravitons are the same as the usual ones but they happen
only in the $x^*$-world.

When $T^a$ represents a fixed source, like the Sun, only $T^a_{00}$ is non
zero. If $T^b$ is  associated with a massless particle the only relevant
part of the propagator gives~\cite{gra1}
\begin{equation}
g^{*2}\,T^a_{00}\,T^b_{00}\,
\frac{1}{p^2-i\epsilon}.
\label{15}
\end{equation}
We do not know the value of $c^*$, however $g^*$ must have a value
consistent with the data concerning the bending of a ray passing near the Sun,
or the perihelion of the movement of Mercury.
However, once we have admited that $c^*\not=c$
we can define a new energy scale related to the Planck scale as
 \begin{equation}
E^{\rm new}=\sqrt{\frac{\hbar{c^*}^5}{G_N}}\equiv
\sqrt{\frac{\hbar c^5}{G_N}}\,\left(\frac{c^*}{c}
 \right)^{\frac{5}{2}}\approx 10^{19}\,\left(\frac{c^*}{c}
 \right)^{\frac{5}{2}}\mbox{GeV}.
\label{1}
\end{equation}

A possible implication of our approach is related to one of the main problems
of
the standard cosmological model. This is the difficulty for explaining the
large-scale uniformity of the observed universe~\cite{scm}. As
information cannot propagate faster than a light signal, in
the standard cosmological model there is not enough time for this
uniformity to be created by any physical process. This is usually
called the {\em horizon problem}~\cite{dp,je}.

The most popular way to
solve this problem is by considering the
inflationary model~\cite{scm,guth}. Notwithstanding, in our context
it is possible to solve this problem if $c^*>c$,
by charging the gravitational waves for the transmission of the information
through the universe.

{}From the point of view of classical general relativity, the weak
gravitational
 field equations in the present context are
\begin{equation}
\Box_x\cdot \Box_{x^*}h_{\mu^*\nu^*}(x,x^*)=16\pi G^{\rm eff}_N\,
T_{\mu^*\nu^*}(x,x^*),
\label{nova}
\end{equation}
where we have defined
\begin{equation}
\Box_{x^*}\equiv \nabla^{*2}-\frac{1}{c^{*2}}\frac{\partial^2}{\partial
t^{*2}},
\label{boxdef}
\end{equation}
being $\nabla^{*2}$ the Laplacian operator with derivatives with respect to
the $\vec x^*$ space coordinates, and $\Box_x$ denoting the usual
d'Alembertian with respect to the $x$--world space-time coordinates;
$G^{\rm eff}_N$ is the effective Newton constant defined in Eq.~(\ref{g}).
 On the other hand, the equation of motion of a Dirac fermion is
\begin{equation}
\left(i\gamma^\mu\cdot\partial_\mu-mc^2\right)_{\alpha\beta}
\left(\Box_{x^*}-m^2c^{*4}\right)\psi_{\beta}(x,x^*)=0,
 \label{f}
\end{equation}
$\alpha$ and $\beta$ are spinor indices in the $x$-world. (We have assumed
that $\hbar$ is the same in both worlds.) Similar modifications will appear in
 the case of other particles, scalar and vector ones.
Although this formalism introduce form factors in all the known interactions
it is not necessary to compactify the extra dimensions as was also pointed out
 in Ref.~\cite{locos}.

Thus, the right-hand side of Eq.~(\ref{nova}) involves the scalar
part of all fields, including the photon. Hence, since light is also affected
only through its scalar part, the limiting velocity in Eq.~(\ref{nova}) is,
in general, different from the local physical velocity of light. So
gravitational waves may have a vacuum velocity different from the velocity of
light. Besides the case $c^*=c$, both possibilities
$c^*>c$ and $c^*<c$ are allowed in this context.
{}From Eq.~(\ref{g}) Newton's law can be written as usual with $G^{\rm eff}_N$.
So, although it is possible that $G^{\rm eff}_N\approx G_N$ for all the
situations considered up to now, obtaining the usual gravitational
interactions, in our context it is also possible that $G^{eff}_N$ can change
with the space-time point.

We would like to emphasize that although the gravitational interactions
propagate through a different space-time they are capable of imparting
momentum to the ordinary bodies.
\acknowledgements

We would like to thank the
Con\-se\-lho Na\-cio\-nal de De\-sen\-vol\-vi\-men\-to Cien\-t\'\i
\-fi\-co e Tec\-no\-l\'o\-gi\-co (CNPq)  partial
financial support. I also thank G. E. A. Matsas for some useful discussions.
I am grateful to Professor D. Bryan for calling my attention for his work in
Ref.~\cite{locos}.


\begin{references}
\bibitem{gra1} Van Dam H and Veltman M 1970 Nucl. Phys. {\bf B22} 397
\bibitem{locos} An eight-dimensional space different from Kaluza-Klein-like
theories in the context of non-gravitational interactions were proposed some
years ago by Bryan D 1986 Phys. Rev. D {\bf34} 1184 and recently by Pleitez V
1995 Extra dimensions and color confinement, preprint IFT-P.019/95
(hep-th/9506009).
\bibitem{gra2} Veltman M 1976 in {\sl Methods in Field Theories},
edited by R. Balian and J. Zinn-Justin, North-Holland, and references
therein.
\bibitem{scm} Kolb E W and Turner M S 1988 {\sl The Early Universe: Reprints},
Addison-Wesley, 1988.
\bibitem{dp} Dicke R H and Peebles P J E 1979 in {\sl General
Relativity}, edited by S.W. Hawking and W. Israel, Cambridge
University Press
\bibitem{je} Ellis J 1994  Nuovo Cimento, {\bf107 A} 1091
\bibitem{guth} Guth A 1081  Phys. Rev. D{\bf23} 347
\end{references}
\end{document}